\documentclass{jps-cp}
\usepackage{txfonts} 

\newcommand\beq{\begin{equation}}
\newcommand\eeq{\end{equation}}
\newcommand\beqa{\begin{eqnarray}}
\newcommand\eeqa{\end{eqnarray}}

\newcommand\btau{\mbox{\boldmath$\tau$}}
\newcommand\bgamma{\mbox{\boldmath$\gamma$}}

\newcommand\bp{{\bf p}}
\newcommand\bk{{\bf k}}
\newcommand\br{{\bf r}}
\newcommand\bq{{\bf q}}

\title{Inhomogeneous Chiral Phase in Quark Matter}

\author{Toshitaka \textsc{Tatsumi}$^{1}$}

\inst{$^{1}$Department of Physics, Kyoto University, Kyoto 606-8502, Japan}

\email{tatsumi@ruby.scphys.kyoto-u.ac.jp}

\recdate{October 5, 2016}

\abst{Recent development of inhomogeneous chiral phase in Kyoto group is briefly reviewed. First, the nesting effect of the Fermi surface is emphasized as a key mechanism leading to inhomogeneous chiral phase. After introduction of inhomogeneous chiral phase, some topological aspect is discussed in 1+1 dimensions and in the presence of the magnetic field in 1+3 dimensions.
Spectral asymmetry gives rise to anomalous quark number and is closely related to chiral anomaly. It may induce spontaneous magnetization and a novel Lifshitz point may appear on the line $\mu=0$. Some astronomical implications are briefly discussed. Finally, the effects of the fluctuations around the order parameter are discussed; the propagator of the fluctuations exhibits a singularity at finite momentum and changes the properties of the phase transition through loop diagrams.  
}

\kword{QCD phase diagram, chiral transition, inhomogeneous condensate}

\begin{document}
\maketitle

\section{Introduction}
 
Nowadays many studies has been devoted to understand the QCD phase diagram in the temperature ($T$), density ($\mu$) and magnetic field ($B$) space. Among them deconfinement transition and chiral transition have attracted much attention and extensively studied theoretically or experimentally. In this talk we focus our attention on chiral transition. In the low density or low temperature region chiral symmetry is spontaneously broken (SSB). Many studies have shown that there occurs restoration of chiral symmetry as temperature or density is increased, while there has been performed no lattice QCD calculation due to sign problem. However, it may be natural in the context of the superconducting model of the vacuum by Nambu and Jona-Lasinio; some quenching effects are caused by the Pauli principle or the thermal effect to suppress the creation of $q{\bar q}$ pairs.

Recently, a possible appearance of the inhomogeneous chiral phase (iCP) has been suggested near the chiral transition and extensively studied in various circumstances \cite{bub}. 
iCP is characterized by the generalized order parameter,
\beq
M\equiv \langle {\bar q}q\rangle+i\langle{\bar q}i\gamma_5\tau_3 q\rangle=\Delta(\br){\rm exp}(i\theta(\br))
\eeq
for $SU(2)_L\times SU(2)_R$, where not only scalar condensate but also pseudoscalar condensate is present. The amplitude or phase may be spatially modulating, and the choice, $\Delta={\rm const.}, \theta=0$, corresponds to the usual chiral order parameter. For example, a configuration called {\it dual chiral density wave} (DCDW) takes the following form (Fig.~1.)\cite{dcdw},
\beqa
\langle{\bar q}q\rangle&=&\Delta {\rm cos}(qz),\nonumber\\
\langle{\bar q}i\gamma_5\tau_3 q \rangle&=&\Delta {\rm sin}(qz).
\eeqa 
Order parameter is rotated in the chiral space, $U(1)$ spanned by scalar and pseudoscalar condensates, as the wave proceeds along one spatial direction.  

\begin{figure}
\begin{center}

\includegraphics[scale=0.5]{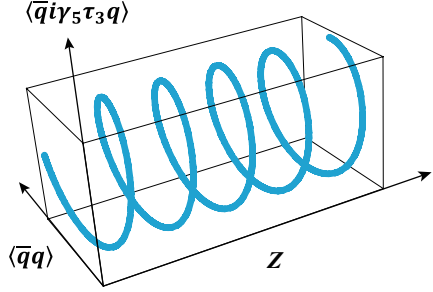}
\end{center}
\caption{Profile of DCDW, which proceeds along $z$ direction.
}
\label{fig1}
\end{figure}
This talk is based on the works in collaboration with S. Karasawa, K. Nishiyama, R. Yoshiike, T-G. Lee and N. Yasutake.

\section{Inhomogeneous Chiral Phase and The Nesting Effect}

There are two popular configurations for iCP, described by the complex quark condensate: one is the plane-wave type of DCDW \cite{dcdw} and the other the standing-wave type called {\it real kink crystal} (RKC) \cite{nic}. Both are the self-consistent solutions of the NJL model within the mean-field approximation and share some common features. iCP is surrounded by the two phase boundaries and the meeting point (Lifshitz point) is the triple point where two uniform phases (SSB and chiral-restored phases) and iCP coexist. Recently, the right -boundary between iCP and the chiral-restored phase has been carefully discussed by taking into account the quantum and thermal fluctuations of the chiral pair fluctuations \cite{kar,yos2}.

How such iCP is brought about? The situation should be clearer for DCDW. We can understand the mechanism in terms of {\it nesting}. The energy spectrum can be given by diagonalizing the Hamiltonian within the mean-field approximation. The spectrum is symmetric with respect to zero and the positive-energy solutions render 
\begin{equation}
\varepsilon^{\pm}({\bf p})=\sqrt{\varepsilon_p^2+|\bq|^2/4\pm\sqrt{(\bq\cdot\bp)^2+m^2|\bq|^2}}
\end{equation}
, using the NJL model in the chiral limit, where $m=-2G\Delta$ and $\varepsilon_p=(p^2+m^2)^{1/2}$. 
Consider $\bp_\perp=0$ for simplicity. It is then reduced to the similar one in  1+1 dimensions. We can graphically construct it by two steps (Fig.~2.). 
First, draw the spectrum without any interaction. We have then two degenerate spectra, and both levels up to $\varepsilon^{\pm}=\mu$ are occupied. Any shift of momentum does not change physics and we have another two spectra, $\varepsilon^\pm=|q/2\pm p_z|$ by the relative momentum difference of $q$. We then find a level crossing at $p_z=0$. After switching on the interaction, the dynamical mass is generated and two degenerate spectra are rearranged to be final spectra. Accordingly, there is opened an energy gap at $p_z=0$. If $2\mu=q$, the energy gap is generated at the Fermi surface. Thus when only the lower levels are occupied, the total energy is necessarily reduced. This is the nesting effect of the Fermi surface and it leads to the appearance of the spatially modulating phase model-independently \cite{sdw,cdw}. It is, however, well-known that nesting is complete only in 1+1 dimensions, while it is incomplete in higher dimensions, since the geometrical shape of the Fermi surface changes with dimension. 
In the case of DCDW in 1+3 dimensions, we have demonstrated by using the NJL model that the DCDW phase develops in the intermediate density region ,$(3-5)\rho_0$ with $\rho_0$ being the the nuclear saturation density, $\rho_0\simeq 0.17$fm$^{-3}$, and the wave-number is large, $q=O(\mu)$, over the DCDW phase \cite{dcdw}. Thus we can say that it is a reminiscence of nesting.

\begin{figure}
\begin{center}
\includegraphics[scale=0.4]{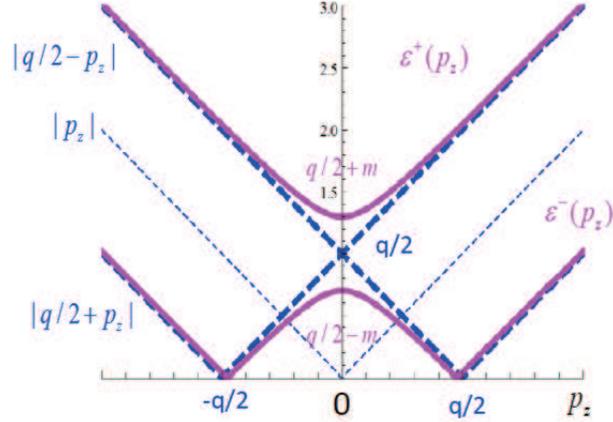}
\end{center}
\caption{Quark energy spectra. Dotted line indicates a free energy spectrum $p_z$ for massless quarks, which is equivalent with $|p_z\pm q/2|$ (dashed lines) 
by a momentum shift of $\pm q/2$. Once the interaction is switched on, the level crossing is resolved to give $\varepsilon^{\pm}(p_z)$. 
}
\label{fig2}
\end{figure}

It should be worth mentioning here that there are two kinds of manifestation of nesting. As is already mentioned, the popular one is the overlapping of the Fermi surface by the shift of the momentum of $O(p_F)$ \cite{sdw,cdw}. There is another type of nesting, which may be relevant to the FFLO state of superconductivity \cite{super}, where two Fermi surfaces of different spin states are no more identical by the external magnetic field. In this case two Fermi surfaces can be overlapped by the small shift of the momentum corresponding to the difference of the Fermi momenta $\Delta p_F$, $q=O(\Delta p_F)$. Thus the wave number starts from zero and smoothly increases in this case by changing the magnetic field. Anyway, an energy gap is opened as well at the Fermi surface. When we consider RKC, we find a similar behaviour as a function of chemical potential. Actually there is {\it duality} between NJL model and a kind of superconducting model in 1+1 dimensions \cite{thi}.

\section{iCP in The Magnetic Field}

Strong magnetic field is nowadays ubiquitous: magnetic field of $O(10^{17})$G may be produced during the high-energy heavy-ion collisions or magnetic field of $O(10^{12-15})$ G have been observed in compact stars. For chiral transition in the presence of the strong magnetic field, enhancement of SSB or tightening of quark-anti-quark pairing have been suggested, which is sometimes called magnetic catalysis\cite{kle,sug,gus}. We'd like to see some topological aspect of iCP, which is manifested in the presence of magnetic field \cite{tatb}.

Recall that chiral spiral is the most favorable configuration in 1+1 dimensions due to chiral anomaly \cite{bas}. Chemical potential $\mu$ can be regarded as a fictitious gauge field $A^\mu=(\mu,0)$ and anomalous Lagrangian reads,
\beq
\delta L=\frac{\theta}{2\pi}\varepsilon^{\mu\nu}F_{\mu\nu}\sim -\frac{\mu q}{2\pi},
\eeq
from which we can find the anomalous number density $\rho_{\rm an}=q/2\pi$. Note that anomalous number density can be also evaluated from the quark energy spectrum,
\beq
\varepsilon_{p,\epsilon}=\epsilon\sqrt{p^2+m^2}+q/2,
\label{spe}
\eeq
which exhibits an asymmetry with respect to the zero line. Atiyah-Patodi-Singer $\eta_H$ invariant is defined by 
\beq
\eta_H={\rm lim}_{s\rightarrow 0}\sum_{\bp,\epsilon}|\varepsilon_{\bp,\epsilon}|^{-s}{\rm sign(\varepsilon_{\bp,\epsilon})},
\eeq
and gives then a measure of spectral asymmetry. Since the quark number operator is given by ${\hat N}=1/2\int dx[\psi^\dagger(x),\psi(x)]$, 
$\eta_H$ is related to number, $N=-\eta_H/2$ at $T=0$\cite{nie}. 
A direct calculation with Eq.~(\ref{spe}) gives $\eta_H=-qL/\pi$. 
Consequently, chiral anomaly gives rise to a surprising result that $q=2\mu$ everywhere in the chiral spiral phase, which implies that nesting is complete in a non-perturbative way. 

Similar situation may be expected in 1+3 dimensions in the presence of magnetic field $B$. The energy spectrum now reads
\beqa
\varepsilon_{n,p,\zeta,\epsilon}&=&\epsilon\sqrt{\left(\zeta\sqrt{m^2+p^2}+q/2\right)^2+2eBn},~~~n\in {\bf Z},\nonumber\\
\varepsilon_{n=0,p,\epsilon}&=&\epsilon\sqrt{m^2+p^2}+q/2,
\eeqa
where $\epsilon=\pm 1$ and $\zeta=\pm 1$ denote particle-anti-particle and spin degrees of freedom, respectively. 
The lowest Landau level ($n=0$, LLL) looks like the one in 1+1 dimensions and exhibits spectral asymmetry, while other levels are symmetric with respect to the zero line (Fig.~3). 
Hence, LLL gives rise to anomalous number,
\beq
\rho_{\rm an}=\frac{eB}{2\pi}\frac{q}{2\pi},
\eeq  
for $q/2<m, \mu<q/2+m$, which is closely related to chiral anomaly \cite{tatb}. The anomalous Lagrangian is then given by 
\beq
\delta L=\frac{\theta}{4\pi^2}F{\tilde F}\sim -\frac{1}{4\pi^2}\mu Bq.
\eeq
Here we'd like to list four important consequences: first, DCDW always appears at $\mu\neq 0$ \cite{fro, nis}. 

Secondly a novel Lifshitz point appears on the line $\mu=0$ on the QCD phase diagram \cite{tatb}. The generalized Ginzburg-Landau expansion of thermodynamic potential renders,
\beqa
\Delta\Omega&=&\frac{\alpha_2}{2}|M|^2+\frac{\alpha_3}{3}{\rm Im}\left(MM'^*\right)+\frac{\alpha_4}{4}\left(|M|^4+|M'|^2\right)+\frac{\alpha_5}{5}{\rm Im}\left(\left(M''-3|M|^2M\right)M'^*\right)\nonumber\\
&&+\frac{\alpha_6}{6}\left(|M|^6+3|M|^2|M'|^2+2|M|^2|M^2|'+\frac{1}{2}|M''|^2\right)+...
\eeqa
near the Lifshitz point within NJL model, where $M'\equiv dM/dz$. The coefficients $\alpha_a$ are the functions of $T,\mu,B$ and $a={\rm odd number}$ are present only for $B\neq 0$. Hence the Lifshitz point emerges 
in the presence of magnetic field only when $\alpha_2=\alpha_3=0$ are satisfied. The latter condition implies $\mu=0$.

Thirdly, we can expect spontaneous magnetization by considering a linear response to the external magnetic field \cite{yos1} (see the next section). Finally, it stabilizes the one dimensional structure to evade the Landau-Peierls theorem \cite{yos}. 


\section{Astronomical Implications}

We briefly discuss some astronomical implications of iCP. First we can see that DCDW provides a new cooling mechanism of hybrid stars 
by supplying additional momentum at the weak-interaction vertex to modify the momentum conservation \cite{mut}.
Considering the quark $\beta$ decay in the DCDW phase, we have estimated the neutrino emissivity as 
\beq
\epsilon_{\rm DCDW}\simeq 6.1\times 10^{26}(\rho_B/\rho_0)^{2/3}Y_e^{2/3}T_9^6~~~({\rm erg\cdot cm^{-3}\cdot s^{-1}}) 
\eeq
near the onset density, where $Y_e$ is the electron number fraction in quark matter, $Y_e=\rho_e/\rho_B$ and $T_9\equiv T/10^9({\rm K})$.
Thus novel cooling mechanism efficiently works and neutrino emissivity becomes almost the same order of magnitude with pion cooling or quark cooling. 
 
Secondly, iCP may provide a scenario for the origin of the strong magnetic field in compact stars \cite{yos1}. As is already mentioned, spontaneous magnetization may emerge in the DCDW phase, depending on the wave number $q$. Considering of the linear response of the DCDW phase to a tiny external magnetic field, spontaneous magnetization is defined by 
\beq
M=-\left.\frac{\partial\Omega}{\partial B}\right|_{B=0}.
\eeq 
In normal phase, thermodynamic potential $\Omega$ is a function of $B^2$, while anomaly gives a linear term with respect to $B$.
Considering a sphere of uniform magnetization $M_0$, the magnetic field , $B=8\pi/3 M_0$ is produced on the surface. In the case of quark matter in the DCDW phase, the magnetic field is estimated to be $10^{16}$G on the surface, which is comparable with the observed value for magnetars.

Finally, a solidification in iCP, especially RKC,  may leads to "chiral crystal" in the high-density core region, in contrast with Coulomb solid or nuclear pastas following the liquid-gas phase transition in the crust region \cite{oka}.
It has a QCD origin of solidification and should open a new possibility for dynamical processes in compact stars; 
it may be the source of oscillation or glitches in hybrid stars. The energy scale should be much greater than that in the crust region in this case.
Unfortunately, there is little studies about higher dimensional configurations in iCP \cite{car}, and should deserve further investigations \cite{yas}.

\section{Fluctuation Effects on The Phase Transition to iCP}

There have been few works about the fluctuation effects beyond the mean-field approximation . 
Two groups have studied the stability of the one-dimensional structure by the Nambu-Goldstone excitations in iCP \cite{lee,hid}. In the case of DCDW two kinds of spontaneous symmetry breaking are induced:
one is the translational or rotational symmetry, and the other is chiral $SU(2)_L\times SU(2)_R$ symmetry. Thus one may expect 4 Nambu-Goldstone modes. However, there is still left one symmetry by combining translation along $z$ direction and chiral $U(1)$ in terms of $Q_5^3$: some shift of phase can be compensated by either transformation. Such NG modes have an anisotropic dispersion, $\omega^2=ak_z^2+bk_\perp^4$. Consequently we can see that the correlation function of the quark bilinear field, ${\bar q}q$ or ${\bar q}i\gamma_5\tau_3 q$, becomes an algebraically  
decaying function due to thermal fluctuations \cite{lee}. This result is consistent with the Landau-Peierls theorem and there is {\it quasi-long-range-order} (QLRO) instead of long-range-order. 
Note that there still exists long range order at $T=0$.

In the recent papers we have studied the effects of fluctuations near the phase boundary of iCP to see change of the properties of the phase transition \cite{kar,yos2}. There are two boundaries which surround iCP in the density-temperature plane. We consider the right boundary which separates iCP and chiral-restored phase, since that boundary shares common features for various configurations.  

We use the two flavor NJL model in the chiral limit.
The partition function reads $Z=\int{\cal D}\psi\int{\cal D}{\bar\psi}e^{-S}$
 with the Euclidean action in imaginary time ($t\rightarrow -i\tau$) being
\beq
S = -\int_0^\beta d\tau\int d^3x \left[
 {\bar\psi}\left(-\gamma^0\frac{\partial}{\partial\tau}+i\bgamma\cdot\nabla+\mu\gamma^0\right)\psi
 + G\left[\left({\bar\psi}\psi\right)^2+\left({\bar\psi}i\gamma_5\btau\psi\right)^2\right] \right],
\eeq
 where $\beta=1/T$ is the inverse temperature and $\mu$ is the chemical potential.
Introducing the auxiliary (collective) fields
 $\phi_a=(-2G{\bar\psi}\psi, -2G{\bar\psi}i\gamma_5\btau\psi)$,
and integrating out the quark field, we have an effective action $S_{\rm eff}$ in terms of $\phi_a$,
\beq
S_{\rm eff}\sim \int d^4x\left[\frac{1}{2!}\Gamma^{(2)}\phi_a^2+\frac{1}{4!}\Gamma^{(4)}(\phi_a^2)^2+...\right],
\eeq
respecting $SU(2)\times SU(2)\simeq O(4)$ symmetry. $\Gamma^{(2)}$ represents the inverse Green function for $\phi_a$ and can be written by the particle-hole pair polarization function $\Pi^0$ in the chiral-restored phase: it reads in the energy-momentum space 
\beq
G(i\omega_n,\bq)^{-1}=1-2G\Pi^0(\bq,i\omega_n)\sim\tau+\gamma(|\bq|^2-q_c^2)+\alpha|\omega_n|
\eeq
near the phase boundary, where $\omega=2n\pi T$ is the Matsubara frequency and $\alpha,\beta,\gamma$ are the functions of $t$ and $\mu$. Note that Green's function have singularities on two dimensional sphere, $|\bq|=q_c$, for $\tau\sim 0$ and $n=0$.
Considering the fluctuations around the thermal average,
\beq
\phi_a=\langle\phi_a\rangle+\xi_a,
\eeq
we can construct the thermodynamic potential in terms of the order parameters $\langle\phi_a\rangle$. 
Since the effective action is chiral symmetric, we can safely choose the thermal average of $\phi_a$ to be $\beta\Phi(\bq)\delta_{n0}=\langle\phi_3\rangle$, 
$\langle\phi_a\rangle=0, a\neq 3$;
\beqa
\Omega-\Omega_f&=&T\sum_q\sum_{\omega_n} {\rm log}\left(1-2G{\bar \Pi}_{\rm ps}^0(q,\omega_n)\right)+\nonumber\\
&+&\frac{1}{2!}
\prod_{i=1}^2 \int\frac{d^3\bq_i}{(2\pi)^3}{\bar \Gamma}_{\rm ps}^{(2)}(\{\bq_i\})\Phi(\bq_1)\Phi(\bq_2) \nonumber \\
&&
 + \frac{1}{4!} 
\prod_{i=1}^4 \int\frac{d^3\bq_i}{(2\pi)^3}{\bar \Gamma}_{\rm ps}^{(4)}(\{\bq_i\})\Phi(\bq_1)\Phi(\bq_2)\Phi(\bq_3)\Phi(\bq_4) \nonumber \\
&&
 + \frac{1}{6!} 
\prod_{i=1}^6 \int\frac{d^3\bq_i}{(2\pi)^3}{\bar \Gamma}_{\rm ps}^{(6)}
 (\{\bq_i\})\Phi(\bq_1)\Phi(\bq_2)\Phi(\bq_3)
 \Phi(\bq_4)\Phi(\bq_5)\Phi(\bq_6) + \cdots. 
\eeqa

\begin{figure}
\begin{center}
\includegraphics[scale=0.4]{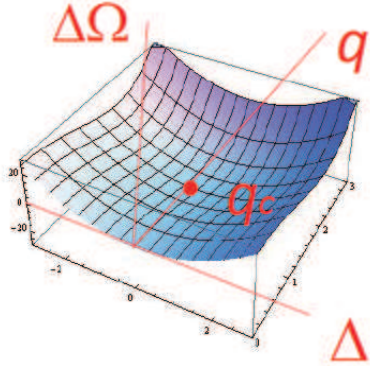}
\end{center}
\caption{Thermodynamic potential $\Delta=\Omega-\Omega_f$ for $\Phi(\bq)=\Delta e^{i\bq\cdot \br}$.
}
\label{fig4}
\end{figure}

The first term represents a contribution of the ring diagrams composed of the polarization function in the chiral-restored phase, 
which is similar to the theory by Nozi\'res and Schmitt-Rink in the case of superconductivity \cite{nsr}. In fact the Thouless criterion holds for the second-order phase transition \cite{tho};
\beq
G(0,\bq_c)^{-1}=0
\eeq
or
\beq
\tau=0, |\bq|=q_c,
\eeq
within the mean-field approximation.
Other terms represent the effective potential $V_{\rm eff}$, where renormalization of the $n$-point vertex function $\Gamma^{(n)}$ is taken into account by the fluctuations. They are consequences of the non-linear effects of the fluctuations and play the role of the effective potential $V_{\rm eff}$ in terms of the order parameter.
We can trace the change of $V_{\rm eff}$ by varying density or temperature, by looking into ${\bar\Gamma}^{(n)}_{\rm ps}$.

The effect of fluctuations changes the two point vertex function $\Gamma^{(2)}$ by modifying the $\tau$ parameter as $\tau_R$. We then find $\tau_R$ never approaches zero by the thermal fluctuations, while quantum fluctuations only shift the critical density. This is due to the dimensional reduction; the effect of singularity in the Green's function becomes more prominent fro the lowest Matsubara frequency at $T \neq 0$.
More importantly, the four point vertex function is affected by 
exchanging two chiral-pair fluctuations (CPF), 
\beq
L(\bk) = T\sum_n\int\frac{d^3\bq}{(2\pi)^3}
 G_{\rm ps}^R(i\omega_n,\bq)G_{\rm ps}^R(-i\omega_n,\bk-\bq),
\eeq
in terms of the renormalized Green's function $G_{\rm ps}^R(i\omega_n,\bq)$ with $\tau_R$, which are called "dangerous terms" and give rise to the longest-range interaction among CPF with momenta, $\bq_1,\bq_2,\bq_3,\bq_4$,
\beq
{\bar\Gamma}^{(4)}_{\rm ps}=(2\pi)^3\lambda\frac{1-\frac{\lambda}{3}L(0)}{1+\lambda L(0)}\delta(\bq_1+\bq_2+\bq_3+\bq_4),
\eeq
where $\Gamma^{(4)}$ is approximated by the contact interaction with the coupling constant $\lambda$.
Since $L(0)$ is positive and diverges as $\tau_R\rightarrow 0$, the second order phase transition is prohibited by CPF at any temperature. This is called fluctuation-induced first order phase transition or Brazovskii-Dyugaev effect \cite{bra,dyu}.

Finally, we briefly discuss some phenomenological implications of CPF. For usual second-order phase transitions, various susceptibilities exhibit anomalous behavior; e.g. specific heat in superconductivity, $c_v=-T\partial^2\Omega/\partial T^2\propto (T-T_c)^{-1/2}$. How about implications of the fluctuation-induced first-order phase transition? We can expect anomalies in the first derivatives in this case, such as entropy or number; e.g. entropy $s$ behaves as $s=s_f-q_cT/2\pi\gamma^{1/2} \tau_R^{-1/2}$ near the phase boundary. Then it should be interesting to get a glimpse of phase transition through relativistic heavy-ion collisions \cite{yos2}.

\section{Summary and Concluding Remarks}

We have seen that inhomogeneous chiral phase (iCP) may be realized in the QCD phase diagram due to the nesting effect of the Fermi surface. Nesting is complete in 1+1 dimensions , where chiral spiral is the most favorite configuration and chiral anomaly plays an important role to give $q=2\mu$. We have shown that DCDW appears by the nesting effect, while it is not complete in 1+3 dimensions. We have also suggested that RKC is also realized by the nesting effect.

DCDW in 1+3 dimensions leads to a topological effect in the presence of the magnetic field. The lowest Landau level exhibits a spectral asymmetry to lead to anomalous quark number. This is closely related to chiral anomaly, where effective electric field is supplied by chemical potential. Note that both DCDW and magnetic field are needed to produce spectral asymmetry. We have seen that such topological effect gives rise to impotant consequences: it remarkably extends the DCDW phase and the Lifshitz point resides on the line $\mu=0$. It also implies spontaneous magnetization in the DCDW phase.

iCP should have some astrophysical implications through cooling mechanism of compact stars or origin of the strong magnetic field in magnetars. Possible solidification (chiral solid) may be also related to some dynamical phenomena in compact stars such as oscillation or glitches. Further investigation should be needed by considering higher dimensional configurations.

We have studied the fluctuation effects on the inhomogeneous chiral transition. Taking into account the chiral pair fluctuations around the condensate in a systematic way, we have found that the order of the phase transition is changed by the fluctuations. It should be interesting to see some continuity of the effect of fluctuations across the phase boundary.
As phenomenological implications we have suggested that anomalous effect can be seen in the first derivatives of thermodynamic potential such as entropy or number density, besides a discontinuous jump of the order parameter.

Finally, we'd like to emphasize a similarity of iCP with other inhomogeneous phase transitions; common features may be hidden in the FFLO state of superconductivity \cite{lee1}.

\end{document}